\documentclass[aps,twocolumn,prl,showpacs]{revtex4-1}
\usepackage{graphicx}
\usepackage{color}
\usepackage{amsmath, amsthm, amssymb}
\usepackage[bf,Large]{}
\begin{document}
\title{Nanoscale phase separation in deep underdoped Bi$_{2}$Sr$_{2}$CuO$_{6+\delta}$ and Ca$_2$CuO$_2$Cl$_2$}
\author{Peter Mistark, Robert S. Markiewicz, Arun Bansil}
\affiliation{Department of Physics, Northeastern University, Boston, Massachusetts 02115, USA}
\date{Version of \today}
\begin{abstract}
We demonstrate that tunneling spectra in deeply underdoped Bi$_{2}$Sr$_{2}$CuO$_{6+\delta}$ (Bi2201) and Ca$_2$CuO$_2$Cl$_2$ (CCOC) provide clear evidence for nanoscale phase separation (NPS), causing the  gap to fill with doping rather than close. The phase separation extends over a doping range from half filling to approximately $x\sim 0.09$.  Assuming the NPS is in the form of stripes, then the nodal gap -- which we model as a Coulomb gap -- arises from impurity pinning of the charged stripes, ultimately driving a metal-insulator transition.
\end{abstract}
\maketitle

Just how a correlated material evolves from a Mott insulator to a high-$T_c$ superconductor remains a highly contentious issue nearly three decades after the discovery of the cuprates, with important implications for the underlying mechanism of superconductivity. A major puzzle in cuprate physics is what happens on doping to the 2~eV-gap present in the half-filled Mott insulator state. Various theoretical models make clear predictions in this connection. In strong coupling theories ($t-J$ model or $U\rightarrow\infty$ Hubbard model) the gap remains large, but there is an anomalous spectral weight transfer (ASWT) from upper (UHB) to lower Hubbard band (LHB).\cite{EMS}  The width of the LHB gradually increases from $\sim 2J$ to $\sim 8t$ as the doping changes from $x=0$ to $x=1$.  For smaller-$U$ Hubbard models, the ASWT is actually faster, as electrons can lower their kinetic energy by hopping through occupied states. In intermediate coupling models, this is associated with a decrease of the magnetic gap with doping.\cite{ASWT}  

However, to explain the rapid changes found at low doping requires accounting for doping-dependent screening of the Hubbard $U$, and in a two-dimensional system with magnetic gap, the screening should turn on discontinuously with doping away from half filling\cite{MKII}.  In such a case, Mott showed that the transition would be first order.\cite{Mott}  This would lead to a very different scenario for the doping dependence of the gap, leading to a filling in rather than a closing as islands of doped phase appear in the sample.  Here we will compare these two scenarios to recent scanning tunneling microscopy (STM) experiments on cuprates.

Experimental data in the deeply underdoped regime, which could help discriminate between different theoretical scenarios, however, have been difficult to obtain until recently. Photoemission sees only filled states, and it is unable to probe the Mott gap. Scanning tunneling microscopy (STM) found only a $\sim 100$~meV pseudogap\cite{Davis_STM}, with no indication as to how this gap connected to the 2~eV optical gap at half-filing. Resonant inelastic x-ray scattering (RIXS) finds evidence for gap collapse, but since it measures a joint density of states (DOS), the analysis is model dependent\cite{MBRIXS}. Very recent STM data from Ca$_2$CuO$_2$Cl$_2$ (CCOC)\cite{CCOC_STM} and Bi$_{2}$Sr$_{2}$CuO$_{6+\delta}$ (Bi2201)\cite{SNS13} give new insight into this problem as these data show the presence of a large gap at half-filling, comparable to the optical gap. Remarkably, the gap in the STM spectra neither remains unchanged nor shrinks with doping, but instead it fills in. This observed behavior is not consistent with a uniform doping scenario. A possible explanation is provided by a model involving competing magnetic orders in which there can be a phase separation between the undoped insulator and an incommensurate magnetic phase near 1/8 doping.\cite{GML} Since the positive ions are fixed in the lattice, this electronic phase separation cannot be macroscopic, but must be a nanoscale phase separation (NPS), possibly in the form of stripes.

\begin{figure} 
\includegraphics[scale=0.17,keepaspectratio=true]{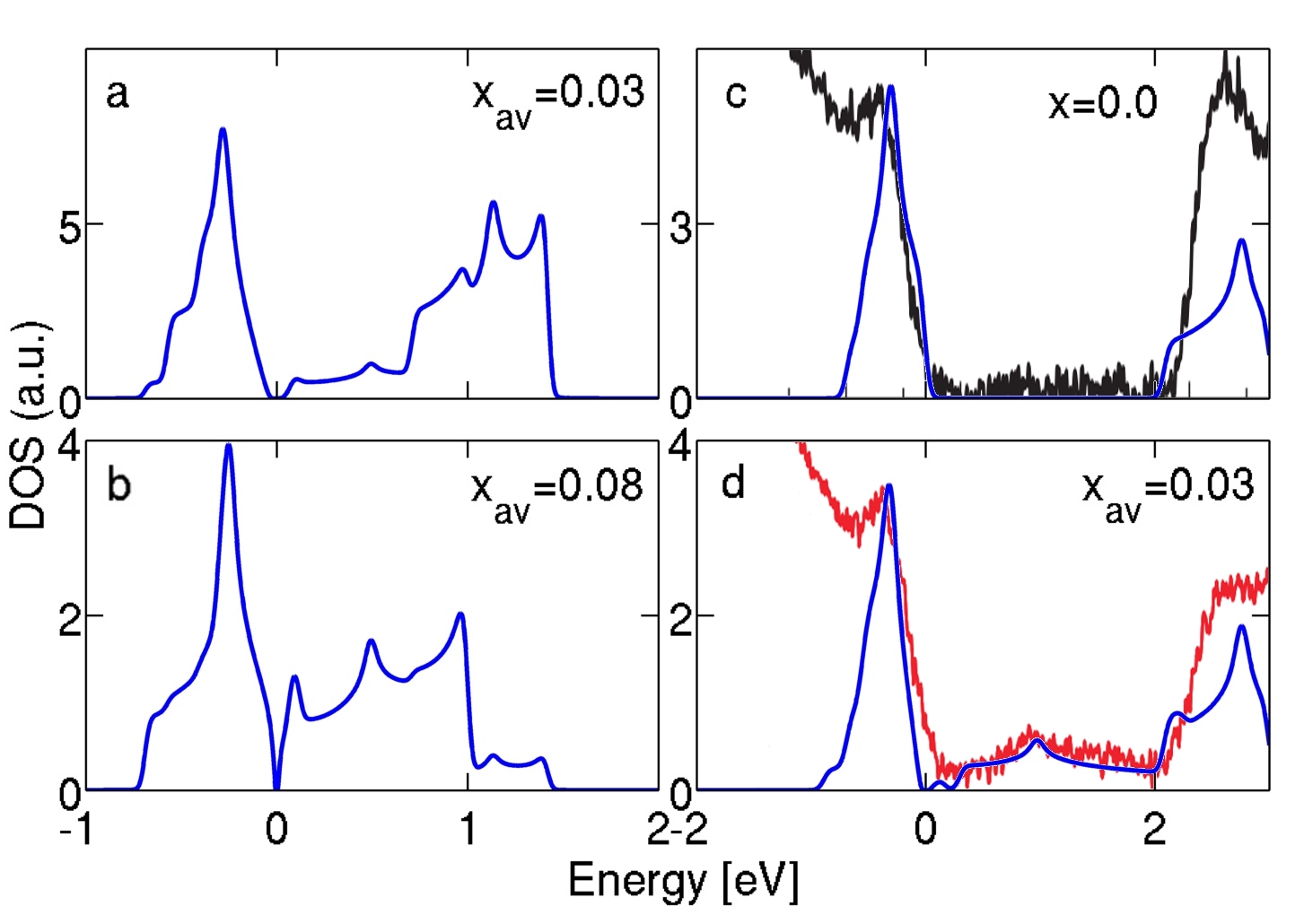}
\caption
{(Color online) LDOS for the AF NPS system, for $x_{av}$ = 0.03 (a) and 0.08 (b) for Bi2201. Parameters used for these curves are $x_0$=0.015, $x_1$=0.09, with a Gaussian broadening of 20meV, and $\Delta$=153.4meV in (a) $\Delta$=20meV in (b). (c) Calculated half filling LDOS (blue) for CCOC with Gaussian broadening of 40meV compared to STM data\cite{CCOC_STM} (black).  (d) Calculated LDOS for the AF NPS system (blue) compared to CCOC data\cite{CCOC_STM} (red). STM data in (c,d) are scaled to match the Van Hove Singularity below the Fermi energy. Here, $x_0$=0.0, $x_1$=0.09, Gaussian broadening of 40meV, and $\Delta$=153.4meV with $x_{av}$ = 0.03. 
\label{fig:3}
}
\end{figure}

Here we show how these results can be understood within an intermediate coupling model.\cite{JoukoDoping}  The band dispersion is taken from density functional theory calculations renormalized by correlations\cite{markietb,ASWT}, while the magnetic order is calculated self consistently within the random phase approximation (RPA).  For undoped Ca$_2$CuO$_2$Cl$_2$ (CCOC) the model predicts a large, $2\Delta\sim$~3~eV gap, consistent with experiment\cite{CCOC_STM}, as well as both the gap and the separation of the two Van Hove singularities (VHSs), Fig.\ref{fig:3}(c).  Moreover, the same model reproduces the experimental dispersions and density of states (DOS) for $x\ge 0.1$, in both Bi$_{2}$Sr$_{2}$CuO$_{6+\delta}$  (Bi2201)\cite{Peter1} and Bi$_{2}$Sr$_{2}$CaCu$_2$O$_{8+\delta}$  (Bi2212)\cite{JoukoDoping}.

\begin{figure} 
\includegraphics[scale=0.17,keepaspectratio=true]{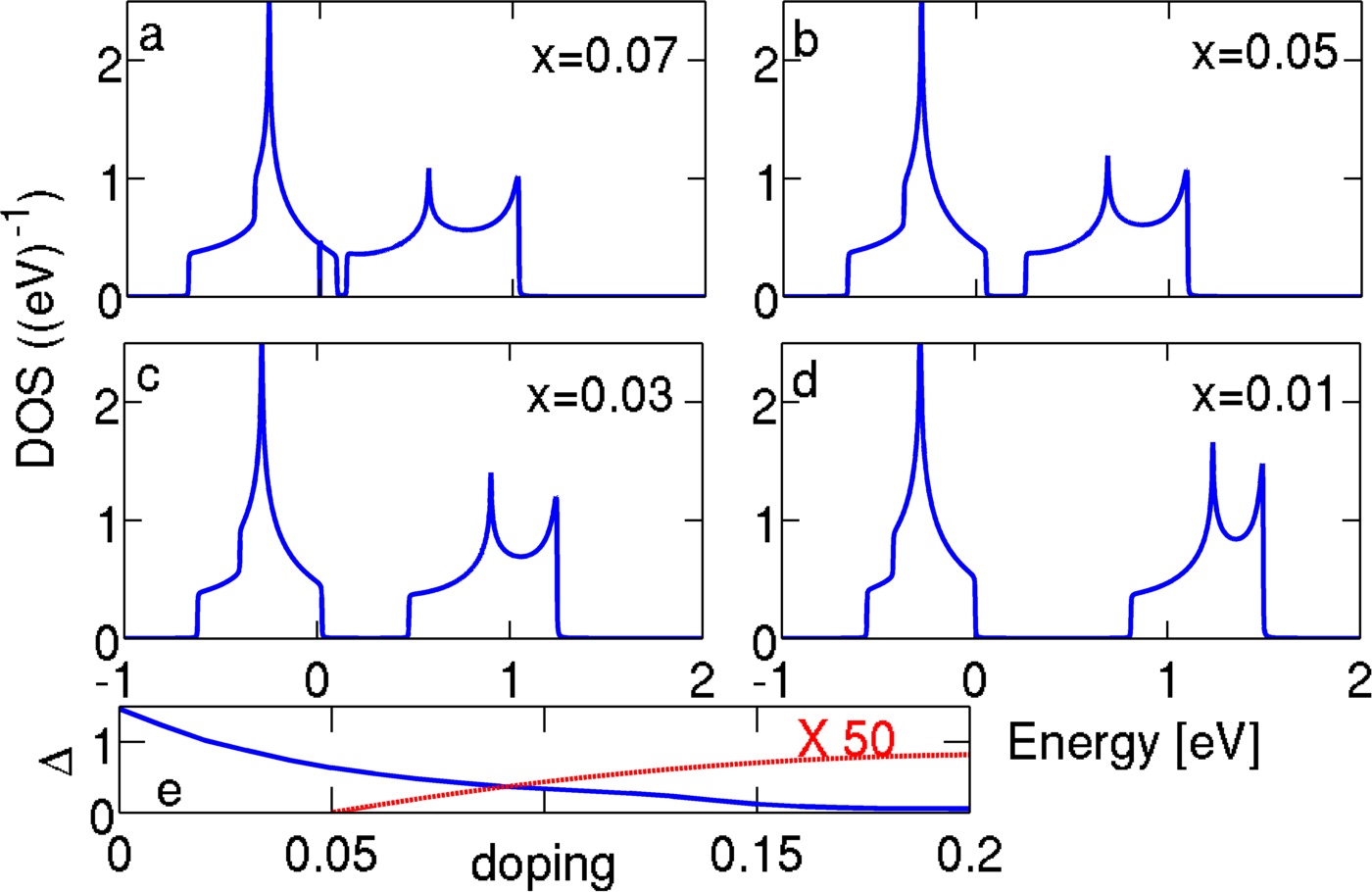}
\caption
{
(Color online) (a-d) LDOS for a uniform antiferromagnetic (AF) plus superconducting (SC) system for Bi2201 at several dopings, $x$ = 0.07 (a), 0.05 (b), 0.03 (c), and 0.01 (d). (e) Two gap scenario, showing AF and SC gaps vs doping.  (The near-FS dip in (a) is due to the superconducting gap.)
\label{fig:2}
}
\end{figure}

However, for a uniformly doped system, the $\sim$~3~eV AF gap at half filling is found to decrease rapidly with doping, leaving only a $\sim$~300~meV pseudogap at $x=0.10$, Fig.~\ref{fig:2}(e).  This would lead to a steady decrease of the gap with increasing doping, as illustrated in Fig.~\ref{fig:2}.  In contrast, recent scanning tunneling microscopy (STM) experiments on deeply underdoped CCOC and Bi2201 reveal a more complicated doping evolution, with a strong growth of  in-gap states in the local DOS (LDOS) spectra.\cite{CCOC_STM,SNS13}  Here we shall show that these results can be understood in terms of nanoscale phase separation (NPS).

In the cuprates, there are several sources of charge inhomogeneity that may act in parallel. First, a charge-density wave phase has been found in a number of cuprates.\cite{EHud,YCDW1,YCDW2,YCDW3,YCDW4,YCDW5,YCDW6}  Then, STM studies find patches of varying local density, which are correlated with oxygen interstitials and vacancies in the Bi-cuprates.\cite{Jenny}  Finally, phase separation has been predicted in Bi-cuprates, most notably between the insulating phase in the undoped cuprate and a doping near $x=0.125$,\cite{GML} essentially the range we are modeling.  As noted above, this leads to NPS.   NPS differs from macroscopic phase separation in that in the latter case only two densities are involved, so that properties such as the AF or superconducting (SC) gaps are doping independent.  In NPS, the individual domains are so small that properties evolve with doping in each domain type, due to proximity to the other domains.\cite{Kapitulnik,RSMstripes}  For instance, in oxygen-doped La$_2$CuO$_{4+\delta}$, the excess oxygen remains mobile to temperatures below room temperature.  In this case, there is macroscopic phase separation, with a wide doping range where the SC transition temperature $T_c$ has a fixed value close to optimal doping, while the AF N\'eel temperature is nearly unchanged from its value at zero doping.\cite{LCO}  In contrast, for Sr-doped La$_{2-x}$Sr$_x$CuO$_{4}$ (LSCO), the $Sr$ ions are immobile, and macroscopic phase separation is replaced by spin-and-charge stripes\cite{KT}, which can be a form of NPS.\cite{VHS00}  Recent neutron scattering studies in Bi2201 find an incommensurate spin response\cite{Tran2201} very similar to the stripe response found in LSCO,\cite{Tranq} suggesting that a very similar NPS is found in both LSCO and Bi2201.


In an NPS model of stripes, the DOS is found to be similar to the superposition of the two separate end phases, and the main effect of nanoscale proximity is to produce small shifts in the density of these end phases -- e.g., the undoped insulator phase will acquire a small hole doping\cite{RSMstripes,foot1}.  We adopt this model here. Figure~\ref{fig:3} displays the LDOS calculated in the NPS model for Bi2201, frames (a) and (b), and CCOC\cite{foot2}, frames (c) and (d). The dopings for CCOC corresponding to the STM data of Ref.~\onlinecite{CCOC_STM} which is plotted along with the calculated curves.  We assume that each patch contains a stripe-like mix of the two phases, so the resulting LDOS is approximately a superposition of the two components, with a weak Gaussian broadening. For Bi2201, Figure~\ref{fig:3}(a,b), we assume that the two stable phases have $x_0$ = 0.015 and $x_1$ = 0.09, and generate linear combinations of the respective LDOSs to represent $x_{av}$ = 0.03, Fig.~\ref{fig:3}(a) and  0.08, Fig.~\ref{fig:3}(b). For CCOC, Figure~\ref{fig:3}(c,d), the stable phases have $x_0$ = 0.0 and $x_1$ = 0.09. This simple picture reproduces the effect of {\it gap filling} as a function of doping, rather than the gap closing shown in Fig.~\ref{fig:2}. In Figs.~\ref{fig:3}(c,d) the model calculations are compared to experimental data (solid lines with noise).  The agreement is quite good -- the model captures both the gap edge and the DOS peak (subband VHS), and, for the doped sample, the in-gap dos.  Note that at energies $<-1$~eV or $>3$~eV there is additional DOS weight associated with bands not included in the theory.




\begin{figure} 
\includegraphics[scale=0.045,keepaspectratio=true]{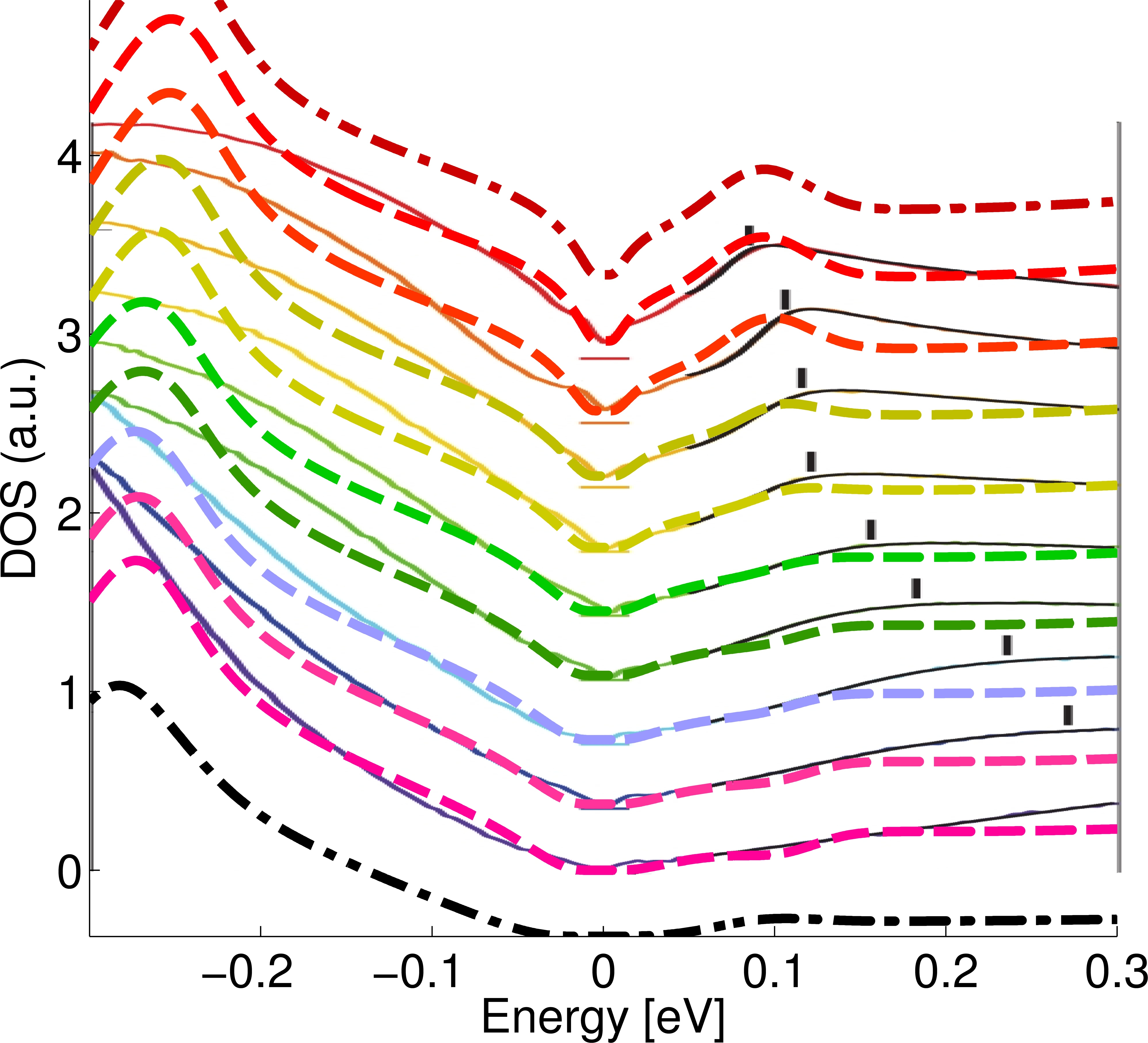}
\caption
{(Color online) calculated LDOS with Gaussian broadening, and Coulomb gap for an AF NPS system (dashed lines) compared to STM data\cite{Davis_STM} (solid lines). The curves are shifted vertically for clarity. The different experimental curves correspond to LDOS taken on different patches in a single Bi2201 sample. The dashed-dotted LDOS curves, at the top and bottom are data for $x_{ave}$ = 0.08 and $x_{ave}$ = 0.03 from Fig. ~\ref{fig:3}(a) and (b) respectively, show that the trend in LDOS will continue to larger and smaller values of $x_{ave}$.  
\label{fig:4}
}
\end{figure}

 To further verify the model, we have also used this two-component model to describe the wider variety of patches found in Ref~\onlinecite{Davis_STM}, but over a narrower voltage range, Fig.~\ref{fig:4}.  We will use this data to describe our calculation in more detail.   To describe the spectra at low voltages, there is one feature we must add to the model.  There is a metal-insulator transition associated with a nodal gap\cite{Kyle,ding,lupi,Davis_STM,Inna}.  Within the present model, this gap arises on the charged stripes (regions of higher doping) and increases as the doping decreases and the stripes separate further.  We assume that it is due to stripe pinning by impurities, and model it as a Coulomb gap\cite{Efros}, as has been suggested previously\cite{ding,Zhou_nodal_gap}. The Coulomb gap is a soft gap in the density of states which is due the the Coulomb interaction of particles on impurity sites. The magnitude of the gap is related to the intensity of the Coulomb interaction. The effect of the Coulomb interaction in two dimensions on the DOS can be calculated self consistently using the following equation derived by Efros\cite{Efros}:
\begin{eqnarray}\label{Cgap}
g(\omega)=\Delta exp\left(-\int_0^{\infty}\frac{g(\omega^{\prime})d\omega^{\prime}}{(\omega+\omega^{\prime})^2}\right),
\end{eqnarray}
where $g(\omega)$ is the resulting DOS at frequency $\omega$ and $\Delta$ is the width of the Coulomb gap. $g(\omega)$ is multiplied by the LDOS calculated using the mean field tight-binding model with AF order to simulate the presence of a Coulomb gap. By applying this Coulomb gap and a Gaussian broadening to LDOS for the AF NPS system our model reproduces the characteristic features of the STM data quite well. Namely, the peak above the Fermi energy due to the bottom of the UMB in $x_1$, the peak below the Fermi energy from the LMB in $x_0$, and the soft gap with zero states at the Fermi energy due to the Coulomb gap. We note that the stable phase, $x_1$, must fluctuate a small amount ($\le$0.015) to account for the shift in energy of the positive energy peak across regions of the sample. To compensate for the fluctuation, $x_0$ is shifted an equal amount such the the difference in doping between $x_1$ and $x_0$ is 0.075. Table I lists the parameters used for the calculated, dashed curves in Fig.~\ref{fig:4} from the top down. The Gaussian broadening in energy has a width of $20meV$ for all calculations.

\begin{table}[h]
\caption {Parameters for Fig.~\ref{fig:4} calculated LDOS} \label{tab:title}
\begin{tabular}{||c|c|c|c|c||}
               \hline\hline
LDOS&$x_0$&$x_1$&$\Delta (meV)$&$x_{av}$ \\
       \hline\hline
1&0.015&0.09&20&0.07875  \\     \hline  
2&0.0125&0.0875&28.3&0.074  \\     \hline
3&0.007&0.082&36.7&0.067  \\     \hline
4&0.005&0.08&45&0.06125  \\     \hline
5&0.004&0.079&53&0.0565  \\     \hline
6&0.003&0.078&61.7&0.05175  \\     \hline
7&0.002&0.077&70&0.047  \\     \hline
8&0.001&0.076&78.3&0.04225  \\     \hline
9&0.0&0.075&86.7&0.0375  \\     \hline
\end{tabular}
\end{table}

The role of NPS or `stripe' physics near a Mott transition has been discussed often\cite{KT,Nag,VHS00}. We note that NPS bears a resemblance to the strong-coupling effect of anomalous spectral weight transfer (ASWT).\cite{EMS}  ASWT is generally interpreted in terms of Mott physics: there is a penalty $U$ for putting a second electron on a copper site that is already occupied.  Thus, when an electron is removed from a given site, both holes lie at a low energy above the Fermi energy, as there is no $U$-penalty for adding a hole with either spin.  Thus, because of ASWT the occupation of the upper Hubbard band is not fixed, but decreases with increased hole doping.
In an intermediate coupling model, the Mott gap becomes an AF gap.  In the presence of NPS, adding a hole creates a region of higher doping, where the AF gap is considerably smaller, so the second hole is shifted to a much lower energy -- e.g., as in Fig.~\ref{fig:3} -- just as in ASWT.  Finally, in the strong coupling regime there is a tendency for atoms with two holes to cluster, to increase the kinetic energy without introducing a $U$-penalty, thus providing an additional link with NPS.

In conclusion, we have demonstrated that the gap filling -- rather than gap closing -- found in STM studies of extremely underdoped Bi2201 is most naturally understood in terms of NPS, as had been predicted in this doping range\cite{GML}. 
Local stripes would be strongly pinned by impurities, naturally explaining the occurance of a nodal gap, a metal-insulator transition, and spin-glass-like phenomena found in underdoped cuprates.  The model further predicts\cite{GML} the coexistence of $(\pi,\pi)$ AF order and an incommensurate SDW phase, as has recently been observed in the closely related compound, LSCO.\cite{AFSDW}.

{\bf Acknowledgments} This research is supported by the US Department of Energy, Office of Science, Basic Energy Sciences contract DE-FG02-07ER46352, and benefited from Northeastern University's Advanced Scientific Computation Center (ASCC), theory support at the Advanced Light Source, Berkeley, and the allocation of supercomputer time at NERSC through grant number DE-AC02-05CH11231. 


\begin{thebibliography}{99}

\bibitem{EMS} H. Eskes, M.B. Meinders, and G.A. Sawatzky, Phys. Rev. Lett.  {\bf 67}, 1035 (1991).
\bibitem{ASWT}R.S. Markiewicz, Tanmoy Das, and A. Bansil, Phys. Rev. B {\bf 82}, 224501 (2010).
\bibitem{MKII}R.S. Markiewicz, Phys. Rev. B {\bf 70}, 174518 (2004).
\bibitem{Mott}N.F. Mott, Phil. Mag B{\bf 50}, 161 (1984).
\bibitem{Davis_STM}Y. Kohsaka, T. Hanaguri, M. Azuma, M. Takano, J. C. Davis, H. Takagi. Nature Physics {\bf 8}, 543 (2012).
\bibitem{MBRIXS}R.S. Markiewicz and A. Bansil, Phys. Rev. Lett. {\bf 96}, 107005 (2006).
\bibitem{CCOC_STM}Cun Ye, Peng Cai, Runze Yu, Xiaodong Zhou, Wei Ruan, Qingqing Liu, Changqing Jin, and Yayu Wang,
Nature Communications {\bf 4}, 1365 (2013). 
\bibitem{SNS13}Yayu Wang, SNS 2013.
\bibitem{GML}G. Seibold, R.S. Markiewicz, and J. Lorenzana, Phys. Rev. B {\bf 83}, 205108 (2011).
\bibitem{JoukoDoping}J. Nieminen, I. Suominen, T. Das, R. S. Markiewicz, and A. Bansil. Phys. Rev. B. {\bf 85}, 214504 (2012).
\bibitem{markietb} R.S. Markiewicz, S. Sahrakorpi, M. Lindroos, H. Lin, and A. Bansil, Phys. Rev. B {\bf 72}, 054519 (2005).
\bibitem{Peter1}P. Mistark, H. Hasnain, R. S. Markiewicz, and A. Bansil. arXiv:1403.2316
\bibitem{EHud}W.D. Wise, Kamalesh Chatterjee, M.C. Boyer, Takeshi Kondo, T. Takeuchi, H. Ikuta, Zhijun Xu, Jinsheng Wen, G.D. Gu, Yayu Wang, and E.W. Hudson, Nature Physics {\bf 5}, 213 (2009).
\bibitem{YCDW1}T. Wu, H. Mayaffre, S. Kr\"amer, M. Horvati\'c, C. Berthier, W.N. Hardy, Ruixing Liang, D.A. Bonn, and 
M.-H. Julien, Nature {\bf 477}, 191 (2011).
\bibitem{YCDW2}G. Ghiringhelli, M. Le Tacon, M. Minola, S. Blanco-Canosa, C. Mazzoli, N. B. Brookes, G.M. De Luca, A. Frano, D.G. Hawthorn, F. He, T. Loew, M. Moretti Sala, D.C. Peets, M. Salluzzo, E. Schierle, R. Sutarto, G.A. Sawatzky, 
E. Weschke, B. Keimer, and L. Braicovich, Science {\bf 337}, 821 (2012). 
\bibitem{YCDW3}A.J. Achkar, R. Sutarto, X. Mao, F. He, A. Frano, S. Blanco-Canosa, M. Le Tacon, G. Ghiringhelli, L. Braicovich, M. Minola, M. Moretti Sala, C. Mazzoli, Ruixing Liang, D.A. Bonn, W.N. Hardy, B. Keimer, G.A. Sawatzky, and D.G. Hawthorn, Phys. Rev. Lett. {\bf 109}, 167001 (2012).
\bibitem{YCDW4}J. Chang, E. Blackburn, A.T. Holmes, N.B. Christensen, J. Larsen, J. Mesot, Ruixing Liang, D.A. Bonn, W.N. Hardy, A. Watenphul, M. v. Zimmermann, E.M. Forgan, and S.M. Hayden, Nature Physics {\bf 8}, 871 (2012).
\bibitem{YCDW5}D. LeBoeuf, S. Kr\"amer, W.N. Hardy, Ruixing Liang, D.A. Bonn, and C. Proust, Nature Physics {\bf 9},79 (2013), 
\bibitem{YCDW6}E. Blackburn, J. Chang, M. Hucker, A.T. Holmes, N.B. Christensen, Ruixing Liang, D.A. Bonn, W N. Hardy, M. v. Zimmermann, E.M. Forgan, and S.M. Hayden, Phys. Rev. Lett. {\bf 110}, 137004 (2013). 
\bibitem{Jenny}I. Zeljkovic, Z. Xu, J. Wen, G. Gu,  R. S. Markiewicz, and J. E. Hoffman, Science { 337}, 320 (2012).
\bibitem{Kapitulnik}A.C. Fang, L. Capriotti, D.J. Scalapino, S.A. Kivelson, N. Kaneko, M. Greven, and A. Kapitulnik, Phys. Rev. Lett. {\bf 96}, 017007 (2006).
\bibitem{RSMstripes}R.S. Markiewicz. Phys. Rev. B. {\bf 62}, 1252 (2000).
\bibitem{LCO}J.D. Jorgensen, B. Dabrowski, S. Pei, D.G. Hinks, L. Soderholm, B.
Morosin, J.E. Schirber, E.L. Venturini, and D.S. Ginley, Phys. Rev. B{\bf 38},
11337 (1988); M.S. Hundley, J.D. Thompson, S.-W. Cheong, Z. Fisk, and J.E.
Schirber, Phys. Rev. B{\bf 41}, 4062 (1990); P.C. Hammel, A.P. Reyes, Z. Fisk,
M. Takigawa, J.D. Thompson, R.H. Heffner, S.-W. Cheong, and J.E. Schirber, Phys.
Rev. B {\bf 42}, 6781 (1990).

\bibitem{KT} S.A. Kivelson, I.P. Bindloss, E. Fradkin, V. Oganesyan, J.M. Tranquada, A. 
Kapitulnik, and C. Howald, Rev. Mod. Phys. {\bf 75}, 1201 (2003).
\bibitem{VHS00}R.S. Markiewicz, J. Phys. Chem. Solids {\bf 58}, 1179  (1997).
\bibitem{Tran2201}M. Enoki, M. Fujita, T. Nishizaki, S. Iikubo, D.K. Singh, S. Chang, J.M. Tranquada, and K. Yamada, Phys. Rev. Lett. {\bf 110}, 017004, (2013).
\bibitem{Tranq}J.M. Tranquada, 
B.J. Sternlieb, J.D. Axe, Y. Nakamura, and S. Uchida, 
Nature {\bf 375}, 561 (1995); 
J.M. Tranquada, J.D. Axe, N. Ichikawa, A.R. Moodenbaugh, Y. Nakamura, and S. Uchida, 
Phys. Rev. Lett {\bf 78}, 338 (1997).
\bibitem{foot1}There should also be minigaps, but we assume these are washed out by fluctuations and disorder.
\bibitem{foot2}Tight binding parameters for CCOC were approximated by the tight binding parameters for Bi2212 found in [\onlinecite{JoukoDoping}].
\bibitem{Kyle}K.M. Shen, T. Yoshida, D.H. Lu, F. Ronning, N.P. Armitage, W.S. Lee, X.J. Zhou, A. Damascelli, D.L. Feng, N.J.C. Ingle, H. Eisaki, Y. Kohsaka, H. Takagi, T. Kakeshita, S. Uchida, P.K. Mang, M. Greven, Y. Onose, Y. Taguchi, Y. Tokura, Seiki Komiya, Yoichi Ando, M. Azuma, M. Takano, A. Fujimori, and Z.-X. Shen, Phys. Rev. B {\bf 69}, 054503 (2003). 
\bibitem{ding} Z. -H. Pan, P. Richard, Y. -M. Xu, M. Neupane, P. Bishay, A. V. Fedorov, H. -Q. Luo, L. Fang, H. -H. Wen, Z. Wang, and H. Ding, Phys. Rev. B {\bf 79}, 092507 (2009).
\bibitem{lupi} S. Lupi, D. Nicoletti, O. Limaj, L. Baldassarre, M. Ortolani, S. Ono, Yoichi Ando, and P. Calvani, Phys. Rev. Lett. {\bf 102}, 206409 (2009).
\bibitem{Inna}I.M. Vishik, M Hashimoto, R.-H. He, W.S. Lee, F. Schmitt, D.H. Lu, R.G. Moore, C. Zhang, W. Meevasana, T. Sasagawa, S. Uchida, K. Fujita, S. Ishida, M. Ishikado, Y. Yoshida, H. Eisaki, Z. Hussain, T.P. Devereaux, and Z.-X. Shen, 
PNAS {\bf 109}, 18332 (2012).
\bibitem{Zhou_nodal_gap} Yingying Peng, Jianqiao Meng, Daixiang Mou, Junfeng He, Lin Zhao, Yue Wu, Guodong Liu, Xiaoli Dong, Shaolong He, Jun Zhang, Xiaoyang Wang, Qinjun Peng, Zhimin Wang, Shenjin Zhang, Feng Yang, Chuangtian Chen, Zuyan Xu, T.K. Lee, X.J. Zhou. Nature Communications {\bf 4}, 2459 (2013).
\bibitem{Efros} A. L. Efros. J. Phys. C: Solid State Phys. {\bf 9} 2021 (1976).
 %
\bibitem{Nag}E.L. Nagaev, ``Physics of Magnetic Semiconductors" (Moscow, Mir, 1983). 
\bibitem{AFSDW}G. Drachuck, E. Razzoli, G. Bazalitski, A. Kanigel, C. Niedermayer, M. Shi, and A. Keren, 
Nature Communications {\bf 5}, doi:10.1038/ncomms4390.
\end{thebibliography}
\end{document}